\documentclass{JHEP3}
\preprint{LU-TP 06-26\\
  hep-ph/0608080}
 
\usepackage{cite}
\usepackage{epsfig}
\usepackage{color}
\let\normalcolor\relax
\usepackage{graphics}
\usepackage{axodraw}
\usepackage{inputenc}
\usepackage{xspace}
\inputencoding{latin1}
\renewcommand\email[1]{{\scriptsize\tt\href{mailto:#1}{#1}}}

\newcommand{\ybar}{\ensuremath{\overline{y}}}

\newcommand{\kbar}{\ensuremath{\overline{k}}}
\newcommand{\mbar}{\ensuremath{\overline{m}}}
\newcommand{\bbar}{\ensuremath{\overline{b}}}
\newcommand{\lbar}{\ensuremath{\overline{l}}}
\newcommand{\pbar}{\ensuremath{\overline{p}}}
\newcommand{\bperp}{\ensuremath{\bar{b}_{\perp}}}
\newcommand{\qperp}{\ensuremath{\bar{q}_{\perp}}}
\newcommand{\kperp}{\ensuremath{\bar{k}_{\perp}}}
\newcommand{\pperp}{\ensuremath{\bar{p}_{\perp}}}

\newcommand{\Aborn}{\ensuremath{A_{\rm{Born}}}}

\newcommand{\eqref}[1]{eq.~(\ref{#1})\xspace}

\newcounter{enumct}

\skip\footins = 1\bigskipamount plus 2pt minus 4pt                              

\title{\boldmath Gravitational Scattering in the ADD-model at High and Low Energies }

\author{Gösta Gustafson and Malin Sjödahl\\
  Dept.~of Theoretical Physics,
  S\"olvegatan 14A, S-223 62  Lund, Sweden\\
  E-mail: \email{Gosta.Gustafson@thep.lu.se}
    and \email{Malin.Sjodahl@thep.lu.se}}
  
  \abstract{
Gravitational scattering in the ADD-model is considered at both sub- and 
transplanckian energies using a common formalism.
By keeping a physical cut-off in the KK tower associated with virtual
KK exchange, such as the cut-off implied from a finite
brane width, troublesome divergences are removed from the calculations
in both energy ranges. 
The scattering behavior depends on three different energy 
scales: the fundamental Planck mass, the collision energy and the 
inverse brane width.
The result for energies low compared to the effective cut-off 
(inverse brane width) is a contact-like interaction. 
At high energies the gravitational scattering associated with the 
extra dimensional version of Newton's law is recovered.
 }

\begin{document}
 
\sloppy

\section{Introduction}
\label{sec:intro}

The ADD-model \cite{Arkani-Hamed:1998rs,Arkani-Hamed:1998nn,Antoniadis:1998ig}
is an attempt to solve the hierarchy problem,
by introducing extra dimensions in which only gravity is 
allowed to propagate. For distances smaller than the assumed 
compactification radius, $R$ 
\footnote{Here we use the notations of \cite{Giudice:2001ce}, such that $R$ is the radius and not the circumference.}, 
the gravitational potential will then be altered and has the form

 \begin{eqnarray}
  \label{eq:Ikye3}
  \frac{V(r)}{m_1m_2}
  &=& -G_{N(4)} R^n S_n \,
  \frac{\Gamma(n)}{r^{n+1}}
\end{eqnarray}
where $n$ is the number of extra dimensions, 
$G_{N(4)}$ denotes the ordinary 3+1-dimensional Newton's constant, 
$S_n=2\pi^{n/2}/\Gamma(n/2)$ is the surface of a unit sphere in $n$ dimensions 
and $\Gamma(n)$ is the Euler Gamma function.
This implies that the strength of gravity increases much faster with 
smaller distance as compared with the normal $1/r$ behavior,
and the fundamental Planck scale (related to the mass 
scale where the corresponding de Broglie
wave length equals the black hole radius) is reduced and given by
\begin{equation}
M_D = \frac{1}{(8\pi R^n G_{N(4)})^{\frac{1}{n+2}}}.
\label{eq:MD}
\end{equation}

The presence of strong gravity at distances smaller than the 
compactification radius opens up for the possibility of observing 
gravitational scattering and black hole production at collider experiments and 
in cosmic rays. To eliminate the hierarchy problem, and not only reduce it, 
the new Planck scale should be of the order TeV, and LHC
will be a quantum gravity probing machine.

In order to quantify the amount of gravitational interaction, the theory was 
formulated as a field theory in \cite{Han:1998sg,Giudice:1998ck}. 
As the extra dimensions are compactified, the 
allowed wave numbers (and hence momenta) in these dimensions are quantized
Kaluza--Klein (KK) modes. The KK modes can of course enter both as external
and internal particles in the Feynman diagrams derived from the theory. 
When the KK modes are internal 
(as for elastic gravitational scattering) they have to be summed 
over. The problem is that the sum over KK modes diverges for 2 or more 
extra dimensions,  
 
\begin{equation}
  \label{eq:KKsum}
  \sum_{\lbar} {\frac{1}{-m_{\lbar}^2+k^2} } 
  = S_n R^n \int{\frac{m^{n-1}}{-m^2+k^2}}dm.
\end{equation}
Here $\lbar$ enumerates the allowed KK modes with momenta $m_{\lbar}$ 
in the extra dimensions,  
$m=|m_{\lbar}|$, and $k$ is the exchanged 4-momentum in our normal space. 
(We will for simplicity call this object a propagator, despite the fact that 
the Lorentz structure is not included.)

In the original papers \cite{Han:1998sg,Giudice:1998ck} 
this divergence problem was dealt with by introducing a sharp
cut-off, $M_s$, argued to be of the same order of magnitude 
as the Planck mass, as new physics anyhow is expected to occur at the 
Planck scale. Various mathematical forms of cut-offs have also been discussed
in \cite{Giudice:2003tu}. For $n \ge 3$ and momentum transfers small compared to $M_s$,  
the sum was then estimated to give

\begin{equation}
  \label{eq:prop}
  \sim \frac{1}{n-2} R^n M_s^{n-2}
  \approx  \frac{1}{G_{N(4)}(n-2)}\,\frac{M_s^{n-2}}{M_D^{n+2}}.
\end{equation}
In the Born approximation this would lead to a cross section of 
the form \cite{Atwood:1999qd}

\begin{equation}
  \label{eq:sigma}
  \frac{d\sigma}{dz} \sim 
  \frac{s^3}{(n-2)^2}\left(\frac{M_s^{n-2}}{M_D^{n+2}}\right)^2
  F(\mbox{spin},z)
\end{equation}
where $z$ is cosine of the scattering angle in the center of mass system, 
$\sqrt{s}$ the total cms energy, and $F$ 
a function taking spin dependence into account.

Ordinary gravitational scattering in $3+n$ dimensions would correspond to 
a potential $\propto 1/r^{(n+1)}$, but the scattering given by
\eqref{eq:sigma} has a completely different angular behavior.
 In particular the expected forward peak is totally 
absent. Fourier transforming the amplitude in
\eqref{eq:prop} to position gives a $\delta$-function potential, 
$\propto \delta(\bar{r})$, and the 
corresponding Born approximation cross section in \eqref{eq:sigma} is 
therefore isotropic. Thus it is obvious that
the approximation in \eqref{eq:prop} does not contain the full story of 
gravitational scattering in the ADD-model.

An attempt to solve this problem has been presented by Giudice, Rattazzi, 
and Wells \cite{Giudice:2001ce}. These authors point out two important 
facts: 

\emph{i}) For an 
interaction with a large Born amplitude but a short range, the cross section 
is not determined by the Born term alone. 
Higher order loop corrections reduce the cross section and guarantee that the 
unitarity constraint is obeyed.

\emph{ii}) The constant term in eq.~(\ref{eq:prop}), which represents a 
dominant part of the amplitude in \eqref{eq:KKsum}, corresponds to
a contribution to the cross section from zero impact parameter, 
and should therefore give a negligible contribution to the cross section,
at least when the incoming wave packages do not overlap. Consequently 
the important part of the amplitude in \eqref{eq:KKsum} must in this case be 
the smaller $k$-dependent terms, which have been neglected in \eqref{eq:prop}.

In case the interaction is dominated by small angle scattering
the cross section can be calculated in the eikonal 
approximation, in which the all-loop summation exponentiates 
\cite{Glauber,PhysRev.126.766,PhysRev.153.1523}.
The cross section is then given by
\begin{eqnarray}
  \label{eq:sigmaeikonal1}
  \sigma_{\rm{el}} &=& 
  \int d^2 \bperp \, |(1-e^{i\chi(\bperp)})|^2 \nonumber\\ 
  \sigma_{\rm{tot}} &=& \int d^2 \bperp \, 2 Re(1-e^{i\chi(\bperp)})\\
  \label{eq:eikonal}
  \mathrm{with} \,\ \chi(\bperp)&=& \frac{1}{2 s} 
  \int \frac{d^2 \qperp}{(2 \pi)^2} e^{i \qperp \cdot \bperp}
  A_{\rm{Born}}(\qperp^2).
\end{eqnarray}
Thus, if the absolute value of the eikonal, $\chi$, is small compared to 1, 
we in general expect 
small corrections from the higher order loop contributions, while for large 
$\chi$-values the cross section saturates, and the effective integrand in 
\eqref{eq:sigmaeikonal1} is close to 1. We also note that when $\chi$ is real, 
the scattering is purely elastic. 
In this paper we will focus on elastic 
collisions mediated via (multiple) exchange in the t-channel.

It is also pointed out in \cite{Giudice:2001ce} that in the eikonal limit the 
Born amplitude does not depend on the spin of
the colliding particles, and is therefore universal. Expressed in the 
fundamental Planck mass $M_D$ in \eqref{eq:MD} it is given by 
\cite{Giudice:2001ce}

\begin{equation}
  \label{eq:Aborn}
  A_{\rm{Born}}(k^2)=
  \frac{s^2}{M_D^{n+2}}\int \frac{d^n \bar{m}}{k^2-\bar{m}^2}.
\end{equation}

In \cite{Giudice:2001ce} a divergent part is subtracted from the integral in
\eqref{eq:KKsum} or (\ref{eq:Aborn}) using dimensional regularization. This 
subtracted part 
corresponds to a narrow potential localized at $\bar{r}=0$. Although the
remainder is singular for $n$ equal to an even integer, its Fourier transform
(the eikonal $\chi$ in \eqref{eq:eikonal}) is finite everywhere. 
Assuming the eikonal approximation to be applicable in the 
transplanckian region $s \gg M_D^2$, the authors of \cite{Giudice:2001ce} 
thus obtains a reasonable result, where the gravitational 
scattering cross section grows with energy $\propto (s/M_D^{n+2})^{2/n}$. 
However, we ought to be worried by the fact that the part of the amplitude,
remaining after the subtraction, grows 
for larger momentum transfers, and is largest for backward scattering.
This implies that the conditions for the eikonal approximation are not
satisfied. The formal problems with divergent integrals 
also indicate that this result could be regarded as based 
more on physical intuition
than on a solid theoretical foundation. These uncertainties
also make it difficult to estimate the limit beyond which
the result should be applicable, and how the gravitational scattering 
behaves for lower energies. 

In this paper we want to study in more detail the result of various physical
effects, which can tame the divergences. These effects give
effective cut-offs for high-mass KK modes at some scale (here referred to 
as $M_s$), which does not have to be the same as the Planck scale $M_D$.
Our result does indeed
confirm the relevance of the eikonal approximation and the result in 
\cite{Giudice:2001ce} at very high energies. For lower energies the behavior is
different, wide angle scattering is dominant and the amplitude does not 
exponentiate. Instead the all-loop summation gives a geometric series.
This implies that there
will be a change in the energy dependence, and for lower energies
the cross section varies more rapidly, proportional to 
$\propto s^2 M_s^{2n-2}/M_D^{2n+4}$.

We want to emphasize that in this paper we do not discuss phenomena 
like black hole formation or other nonlinear gravitational effects,
which are expected to modify the final states for very high energies and
central collisions.
For a discussion of such effects we refer to ref. \cite{Giudice:2001ce, Antoniadis:1998ig,Emparan:2000rs, Dimopoulos:2001hw,Giddings:2001bu,Kanti:2004nr,Lonnblad:2005ah,Harris:2003eg}.
We also neglect possible interference with strong and electro-weak effects
and we study reactions for non-identical particles such that KK modes appears
only in the t-channel. Some remarks on s- and u-channels are however made 
in secs.\ \ref{sec:Region3}-\ref{sec:Region5}.

The approach in \cite{Giudice:2001ce} will be discussed in 
more detail in sec.\ \ref{sec:Giu}. In sec.\ \ref{sec:Brane}, we will 
introduce a finite width of the brane, on which 
the standard model particles are assumed to live, and see how this leads 
to a finite amplitude. A similar effect is obtained by assuming that the
position of the brane is not fixed in the extra dimensions 
\cite{Bando:1999di,Kugo:1999mf}. Fluctuations in the brane then result in
a kind of surface tension or "brane tension".
The Born term is discussed in sec.\ \ref{sec:Born} and higher order 
loop corrections in sec.\ \ref{sec:loop}. Here we also study in which 
kinematical regions the Born term dominates, where the
eikonal approximation is valid, and the behavior of the cross section
in regions where the scattering is approximately isotropic.
The results for scattering cross
sections in those different kinematical regions are then presented in 
sec.\ \ref{sec:cross}. 
Finally we will summarize and conclude in sec.\ \ref{sec:Conclusion}.

\section{Problems and divergences}
\label{sec:Giu}

The integral in \eqref{eq:KKsum} or (\ref{eq:Aborn}) 
is divergent for $n\geq 2$ and $n\leq 0$, but converges for 
$n$-values in the intermediate range $0<n<2$. To give a physical meaning 
to the integral 
for $n\geq 2$, a finite result can be obtained by analytic continuation 
from smaller $n$-values, corresponding to a dimensional regularization. 
The resulting amplitude, presented in \cite{Giudice:2001ce}, is given by the 
expression\footnote{We have here inserted a minus sign not present in 
\cite{Giudice:2001ce}.}
\begin{equation}
  \label{eq:Aborn2}
  A_{\rm{Born}}(k^2)
  =-\pi^{\frac{n}{2}}\Gamma\left(1-\frac{n}{2}\right)
  \left(\frac{-k^2}{M_D^2} \right)^{\frac{n}{2}-1}
  \left(\frac{s}{M_D^2} \right)^2.
\end{equation}
We see that this expression 
is finite for odd integers $n$, but singular for 
even $n$, where the $\Gamma$-function has poles.

The result in \eqref{eq:Aborn2} is equivalent to a subtraction of terms,
which are proportional to
$\delta$-functions or derivatives of $\delta$-functions at $\bar{r} = 0$,
and therefore may be  expected to give negligible contributions to the 
cross section. Inserting \eqref{eq:Aborn2} into the two-dimensional 
Fourier transform in \eqref{eq:eikonal}, we see that this
integral is also divergent. It can be given a finite result by
introduction of a convergence factor:
\begin{equation}
  \label{eq:chibc}
  \chi=-\left(\frac{b_c}{b}\right)^n ,\,\,\,\mathrm{with} \,\,
  b_c=\left[ \frac{s(4\pi)^{\frac{n}{2}-1}\Gamma(n/2)}
    {2 M_D^{n+2}}\right]^{1/n}.
\end{equation} 
We note that although the amplitude $A_{\rm{Born}}$ in \eqref{eq:Aborn2} is
singular for even $n$, $\chi$ is finite.
Thus $\chi(b)$ (like the potential $V(r)$, to be discussed below) can 
be analytically continued
to finite values for all $n$-values. (A finite amplitude, which
corresponds to a potential proportional to $1/r^{n+1}$ for $n$ even,
is $\propto (-k^2)^{n/2-1}\ln{(-k^2)}$.)

The result in \eqref{eq:chibc} is a single power 
$\propto 1/b^n$, and the scale factor (or characteristic impact parameter)
$b_c$ is defined so that $|\chi| = 1$ when $b=b_c$. If this expression
is inserted into \eqref{eq:sigmaeikonal1}, we see that the term
quadratic in $\chi$, which is the Born term, dominates the integrand 
for $b>b_c$, where $\chi < 1$, but higher order 
corrections are important in constraining the scattering probability for 
$b<b_c$.

In ref. \cite{Giudice:2001ce} it is assumed that eqs.~(\ref{eq:sigmaeikonal1})
and (\ref{eq:chibc}) should 
give a realistic approximation to gravitational scattering in the 
transplanckian region $s \gg M_D^2$ (apart from special effects like 
black hole formation, which are treated separately).
The net result is then that the total scattering cross section grows with 
energy proportional to $b_c^2$, or equivalently $\propto (s/M_D^{n+2})^{2/n}$,
(cf. \eqref{eq:sigmahigh} below).

The exponentiation in the eikonal approximation in \eqref{eq:sigmaeikonal1} 
follows when the scattering is dominated by small angles 
\cite{Glauber,PhysRev.126.766,PhysRev.153.1523}. The one-loop 
contribution is then dominated by its imaginary part, and the all-loop
summation gives an exponential. 

\FIGURE[t]{%
      \epsfig{file=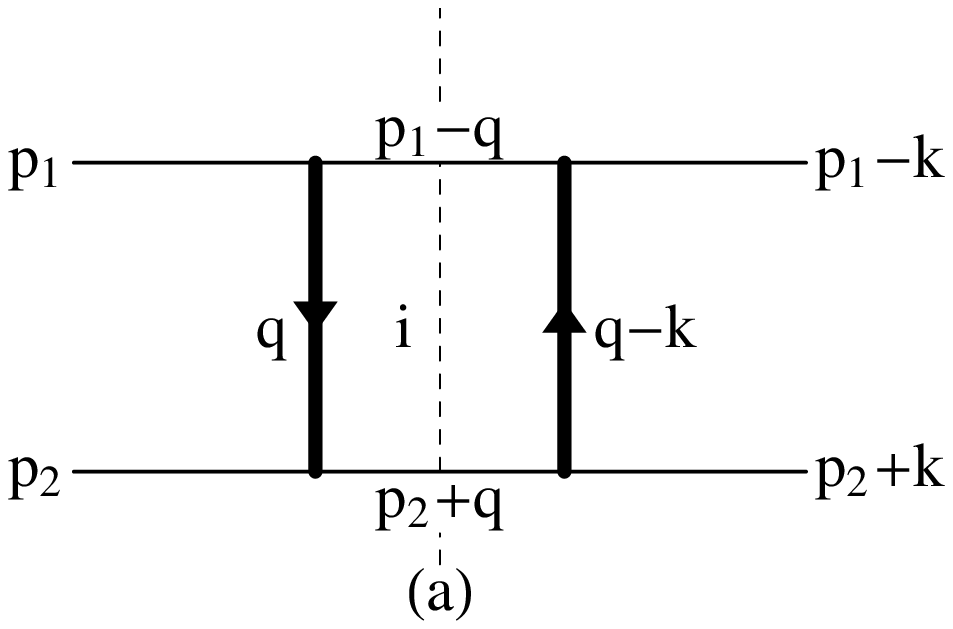,width=7cm}\hfill
      \epsfig{file=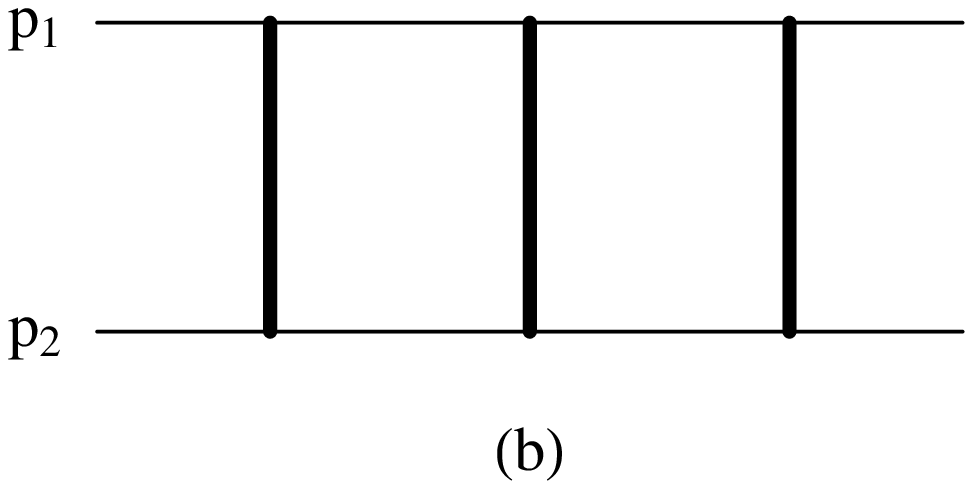,width=7cm}
\caption{\label{fig:Highs} 
(a) The one-loop contribution corresponding to exchange of two KK modes. 
The KK modes are drawn as thick lines and standard model particles as 
thin lines. 
(b) The two-loop contribution.
}}

The one-loop diagram in fig.~\ref{fig:Highs}a is given by the following 
expression:
\begin{equation}
  \label{eq:A1loop}
  A_{\rm{1-loop}}(k^2)=\frac{-i}{2}
   \int\frac{d^4 q}{(2\pi)^4} A_{\rm{Born}}(q^2)A_{\rm{Born}}((k-q)^2)
  \frac{1}{(p_1-q)^2}\frac{1}{(p_2+q)^2}.
\end{equation}
Here $p_1$ and $p_2$ denote the momenta of 
the incoming particles, the total momentum exchange is $k$ and the 
loop momentum $q$. 

The imaginary part of the integral in \eqref{eq:A1loop} is coming from 
on-shell intermediate states (denoted $i$ in fig. \ref{fig:Highs}), 
and can be calculated
using the Cutcosky cutting rules. This implies that the two
propagators in \eqref{eq:A1loop} are replaced by $\delta$-functions,
which (with the approximation $q^2 \approx \qperp^2$) gives the result 
\begin{equation}
  \label{eq:A1loopim}
  A_{\rm{1-loop}}(k^2)=\frac{i}{4s}\int \frac{d^2 \qperp}{(2\pi)^2}
  A_{\rm{Born}}(-\qperp^2)A_{\rm{Born}}(-(\kperp-\qperp)^2).
\end{equation}
If $A_{\rm{Born}}$ falls off for large $k_\perp$, the Fourier
transform to impact parameter space of the one-loop contribution is 
proportional to  $\chi^2(\bperp)$. 
The sum over multi-loop ladder diagrams with different
number of KK exchanges exponentiates to 
$(i\chi-\chi^2/2+...)=e^{i\chi}-1$, and the all order eikonal amplitude 
is given by
\begin{equation}
  \label{eq:Aeik}
  A_{\rm{eik}}(k^2)=-2 i s \int d^2 \bperp e^{i \kperp\cdot \bperp}(e^{i\chi}-1).
\end{equation}

With the Born amplitude in \eqref{eq:Aborn2} we have, however, some 
problems. First the real part of 
the integral in \eqref{eq:A1loop} is not small and negligible compared
to the imaginary part. It is strongly 
divergent for $n\ge 2$, as $A_{\rm{Born}}$ increases proportional to 
$q^{n-2}$ or $q^{n-2}\ln(-q^2)$ for large $q$. Secondly the integral 
in \eqref{eq:A1loopim} should only go over physical intermediate states,
which means $q_\perp < \sqrt{s}/2$. The Fourier transform of the convolution
in \eqref{eq:A1loopim} gives $\chi^2$ only if the integrand in 
\eqref{eq:A1loopim} falls off so rapidly, that the integral effectively can
be extended to all $q_\perp$. This is not the case with the amplitude
in \eqref{eq:Aborn2}.

We conclude that, although the result in \eqref{eq:chibc} and 
(\ref{eq:sigmaeikonal1}) is an intuitively reasonable result
for scattering in a rapidly falling potential, it should
be worrying that it is derived from an amplitude, which grows for large
momentum transfers and large scattering angles, while the eikonal 
approximation is proven to be valid only when scattering at small angles is 
dominating. 

At the root of this problem lies the fact that 
the subtraction, which gives the amplitude in 
\eqref{eq:Aborn2} and is a result of the analytic continuation in the
number of extra dimensions, does \emph{not} automatically remove all parts 
corresponding to $\delta$-functions at $\bar{r}=0$. 
The definition of the potential as the Fourier transform of \eqref{eq:Aborn2}
is problematic.
To illustrate this we study the most simple example represented 
by the case $n=3$. In the rest frame we have $k_0=0$ and $k^2=- \bar{k}^2$. 
The integral in \eqref{eq:Aborn} is then proportional to
\begin{eqnarray}
  \label{delta}
  \int \frac{m^2 \, dm}{\bar{k}^2+m^2} 
  &=& 
  \int \frac{ ( m^2+\bar{k}^2- \bar{k}^2 ) dm } {\bar{k}^2+m^2}
  =
  \int dm - \bar{k}^2 \int \frac{ dm}{\bar{k}^2+m^2} \nonumber \\
  &=&
  \int dm - |\bar{k}| \int \frac{ dx}{1+x^2}=
  \int dm - |\bar{k}| \frac{\pi}{2}.
\end{eqnarray}
The first term, the integral, represents an infinite subtraction. 
Its three-dimensional
Fourier transform gives a $\delta$-function at $\bar{r}=0$ with an 
infinite weight. The second term corresponds to the result in 
\eqref{eq:Aborn2}. 
We may try to define its Fourier transform $\hat{V}(\bar{r})$ as a distribution
in the standard way, multiplying with a test function and interchanging
the order of integration. For test functions of the form $\exp(-a \,r^2)$
we then get (with $k \equiv |\bar{k}|$ and the constant $C$ appearing in 
\eqref{eq:Vbigr})

\begin{eqnarray}
  \int d^3 r e^{-a\, r^2} \hat{V}(\bar{r})
  &\equiv& C (-\frac{\pi}{2}) \int d^3 k\, k \int d^3 r e^{-a\, r^2} 
  e^{i\bar{r} \bar{k}} \nonumber \\
  &=&
  -C\frac{\pi}{2} \int d^3 k\, k e^{-k^2/4a} 
  (\frac{\pi}{a})^{\frac{3}{2}} \nonumber \\
  &=& - C 16 \pi^{\frac{7}{2}} \sqrt{a}.
  \label{eq:Vtrans}
\end{eqnarray}
We note that this result is finite, and goes towards 0 when the test
function approaches a constant, i.e. when $a \rightarrow 0$. For 
$r \neq 0$ we find $\hat{V}(r) =C 4 \pi^2/r^4$ by Fourier transforming from 
$\bar{k}$ to  $\bar{r}$ using a convergence factor. 
Integrating this contribution with the test function above, we get the 
divergent result
\begin{equation}
  C 16 \pi^3 \int_0^\infty e^{-a \, r^2} \frac{d r}{r^2}.
  \label{eq:Vint}
\end{equation}
Thus this definition, $\hat{V}(r) =C 4 \pi^2/r^4$ for $r>0$, is incomplete 
since the result in \eqref{eq:Vtrans} is finite while the integral
in \eqref{eq:Vint} is infinite. It looks as if a 
$\delta$-function, $\delta(\bar{r})$, with infinite weight is missing.

We conclude that the separation in \eqref{delta} does not in itself remove 
all terms related to $\delta$-functions at $\bar{r} = 0$. 
Instead we argue in the next section that dynamical effects will 
remove the divergencies and give finite results.

In the next section we will consider possible
mechanisms, which can suppress high KK masses and give an effective 
cut-off to the integral in \eqref{eq:Aborn}. These mechanisms have
real dynamical motivations, and we will see that such finite cut-offs
do remove all divergences and give well-defined 
results. For high energies the Born amplitude indeed falls off for large
momentum transfers, and the eikonal approximation is applicable. For lower 
energies this is not the case, and we will in secs. \ref{sec:Born} - 
\ref{sec:cross} discuss the resulting amplitudes and cross sections for
different relations between the energy, the Planck mass, and the cut-off scale.

\section{Possible solutions}
\label{sec:Brane}

In the ADD-model the 
standard model particles are assumed to live on a thin 
brane. The mechanism behind this assumption could possibly be taken from 
string theory \cite{Antoniadis:1998ig}, but is not  a part of the 
ADD-model itself. 
The problems discussed in the previous section are related to
contributions from KK modes with very high masses. 
In a relativistic quantized theory there are also formal
problems with an infinitely thin and infinitely rigid brane. 
If the brane is not infinitely thin, but
has a finite width, this will effectively
suppress the coupling to high-mass KK modes, with wavelengths shorter
than the brane width. If the brane really is 
infinitely thin, then it must be impossible to determine its position with
infinite accuracy. In \cite{Bando:1999di,Kugo:1999mf} it 
is demonstrated that the fluctuations in the position
of the brane suppresses high-mass KK modes, in a way similar
to the effect of a finite brane width. The emission or absorption 
of a KK mode gives a recoil to the brane, and the fluctuations in the 
location of the brane can then be regarded as a result of 
an effective "surface tension" in the brane.

Let us for definiteness study the effect of the assumption that the 
standard model fields
penetrate a finite distance into the extra dimensions, which gives an 
effective finite width to the brane \cite{Sjodahl:2006vq}. 
(The possibility of fluctuating branes, studied in 
\cite{Bando:1999di,Kugo:1999mf}, give similar results, 
albeit with a different physical interpretation.) To be specific we assume a 
Gaussian extension, but this assumption is not essential for our conclusions.
Thus we assume the standard model fields to have a wave function
with the extension

\begin{equation}
  \label{eq:SMydist}
  \psi(\ybar)=\left( \frac{M_s} {\sqrt{2\pi}} \right)^{ \frac{n}{2} }  
  e^{-\ybar^2 M_s^2/4}
\end{equation} 
into the extra dimensions, with $\ybar$ denoting the coordinate in the 
extra dimensions. The overlap between two standard
model fields and a KK mode of mass $m$ (what we have in a vertex) 
is then proportional to

\begin{equation}
  \label{eq:overlap}
  \int d \ybar\, e^{i \mbar \cdot \ybar} 
  \left( \frac{M_s}{\sqrt{2 \pi}} \right)^{2\frac{n}{2}}
  e^{-\ybar^2 M_s^2/2} 
  = e^{-m^2/(2 M_s^2)}
\end{equation} 
or, in other words, the squared absolute value 
of the wave function in $\ybar$-space Fourier transformed to $m$-space. 
The exchange of a KK mode will have this suppression factor
occurring twice, once at every vertex. In total the exchange of a KK mode 
with mass $m$ will therefore contribute to the sum in \eqref{eq:Aborn} 
with a suppression factor

\begin{equation}
  \label{eq:sup}
  e^{-m^2/M_s^2}.
\end{equation} 

Implementing the physical requirement that the standard model particles 
live on a narrow brane does therefore in itself imply a finite ``effective''
propagator,
\begin{equation}
  \label{eq:SMprop}
  R^n S_n\int \frac{dm \,m^{n-1}}{k^2-m^2} e^{-m^2/M_s^2}
\end{equation} 
for the exchange of 4-momentum $k$. (The factor $R^n$ comes 
from the density of KK modes and $S_n = 2 \pi^{n/2} / \Gamma(n/2)$ 
is the unit surface of a sphere in $n$ dimensions.) 
We note in particular that this expression (in contrast to the expression in 
\eqref{eq:Aborn2}) falls off like $1/k^2$ for large momentum transfers, 
such that $-k^2 \gg M_s^2$. 
This implies that for high energies, $s \gg M_s^2$,  t-channel 
interaction is dominated by small values of $-k^2/s$, 
i.e. by small angles.

In the following sections we will show that the Fourier transform of
the propagator in \eqref{eq:SMprop} gives a potential, which falls off
$\propto 1/r^{n+1}$ for distances larger than the brane width, given by
$1/M_s$, and smaller than the compactification radius.
Outside this range, both for $r<1/M_s$ and for $r>2 \pi R$ (where the massless
graviton dominates), it varies $\propto 1/r$. 
We will also study the resulting scattering cross 
sections under different kinematic conditions.

\section{The Born term}
\label{sec:Born}

\subsection{Amplitude}
As described in section \ref{sec:Brane}, several physical mechanisms
result in effective cut-offs for high masses in the KK propagator. 
After multiplying \eqref{eq:SMprop} by the coupling 
$4\pi G_{N(4)}$, contracting Lorentz indices (not explicitly included here), 
and using the relation 
$G_{N(4)}^{-1}=8 \pi R^n M_D^{2+n}$ we get the following result 
for the Born amplitude for ultra-relativistic small angle scattering:

\begin{equation}
  A_{\rm{Born}}(t) 
  = \frac{s^2}{M_D^{n+2}}\, S_n \int_{0}^{\infty} 
  \frac{dm \,m^{n-1}}{k^2-m^2}\, e^{-m^2/M_s^2}.
  \label{eq:ABorn}
\end{equation}
For large angels there are less important corrections from spin polarization
which we neglect here and in the following.
This integral is convergent and finite for all negative 
values of $k^2=t$ (including 0 when $n \geq 3$). It is easy to find the result 
in the limits of large and small (negative) $t$-values. 

\begin{itemize}

\item {\emph{Large momentum transfers; $-t \gg M_s^2$}

When $-t$ is large compared to $M_s^2$, the term $m^2$ in the 
denominator in \eqref{eq:ABorn} can be neglected, which gives the result:
\begin{equation}
  A_{\rm{Born}}(t) \approx  \frac{s^2}{M_D^{n+2}}\, S_n \int_0^\infty 
  \frac{dm \,m^{n-1}}{t} e^{-m^2/M_s^2}
  = \pi^{n/2} \left(\frac{M_s}{M_D}\right)^n \frac{s^2}{M_D^2 \cdot t}.
  \label{eq:ABornbigt}
\end{equation}

Thus for large momentum transfers (larger than the cut-off) the Born amplitude
falls off proportional to $1/t$.
}
\item {\emph{Small momentum transfers; $-t \ll M_s^2$}

For smaller $t$, and $n>2$, the integral is dominated by $m$-values of 
the order of $M_s$, and therefore $t$ can now be neglected in the 
denominator. We then get the approximately constant result:

\begin{eqnarray}
  A_{\rm{Born}}(t) 
  &\approx&  
  \frac{-s^2}{M_D^{n+2}}\, S_n \int_0^{\infty} 
  dm \,m^{n-3}\, e^{-m^2/M_s^2} \nonumber \\
  &=&
- \frac{2\pi^{n/2}}{(n-2)}   \left(\frac{M_s}{M_D}\right)^n
\frac{s^2}{M_D^2 M_s^2}. 
  \label{eq:ABornsmallt}
\end{eqnarray}

Thus for momentum transfers, which are small compared to the cut-off, the Born
amplitude is approximately constant for $n>2$. For $n=2$ the result for
small $t$ has instead a slowly varying logarithmic dependence, proportional to 
$\ln(-M_s^2/t)$.
}
\end{itemize}

\subsection{Potential}
To get the classical non-relativistic potential we start directly from the 
effective propagator in \eqref{eq:SMprop} multiplied with the coupling constant
$4 \pi G_{N(4)}$.
Going to the rest frame, where $k_0=0$ and $t=-\bar{k}^2$ we find the
corresponding potential as the three-dimensional Fourier transform:

\begin{eqnarray}
  \frac{V(r)}{m_1 m_2}
  &=& \frac{1}{2 s^2}
  \int \frac{d^3 \kbar}{(2\pi)^3}\, e^{i \bar{k} \bar{r}} 
  A_{\rm{Born}}(-\bar{k}^2)
  \nonumber\\
  &=& 
  \frac{-1}{2M_D^{n+2}} \frac{S_n}{(2\pi)^3} 
  \int_0^\infty dm \,m^{n-1}\, e^{-m^2/M_s^2} 
  \int \frac{d^3 k e^{i \bar{k} \bar{r}}}{m^2 + \bar{k}^2} = \nonumber \\
  &=& \frac{-1}{2M_D^{n+2}} \frac{S_n}{(2\pi)^3} 2 \pi^2 
  \int_0^\infty dm \,m^{n-1} e^{-m^2/M_s^2} \cdot \frac{e^{-mr}}{r}.
  \label{eq:V}
\end{eqnarray}
This represents a weighted sum of Yukawa potentials. The integral can 
be expressed in terms of error functions, but we are here 
primarily interested in the behavior for large and small values of $r$.

\begin{itemize}

\item {\emph{Large distances; $r > 1/M_s$}

For distances larger than the brane thickness the integral 
is effectively cut off by the factor $e^{-mr}$,
and the result becomes insensitive to the Gaussian cut-off $e^{-m^2/M_s^2}$.
It is then approximated by

\begin{equation}
  \frac{V(r)}{m_1 m_2}
  \approx
  \frac{-1}{2M_D^{n+2}} \frac{S_n}{4\pi} 
  \int_0^\infty dm m^{n-1} \cdot \frac{e^{-mr}}{r}
  =
  \frac{-S_n\,\Gamma(n)}{8 \pi M_D^{n+2}}
  \cdot \frac{1}{r^{n+1}}.
  \label{eq:Vbigr}
\end{equation}


We see that for distances large compared to the brane thickness (but
small compared to the compactification radius) we recover the result from
\eqref{eq:Ikye3}, a potential falling off proportional to $1/r^{n+1}$, 
corresponding to 
the expected ($3+n$)-dimensional version of Newton's law. 
When $r$ is increased, smaller $m$-values $\sim 1/r$ 
are important in the integral in \eqref{eq:V} or (\ref{eq:Vbigr}). 
The phase space factor 
$m^{n-1}$ then gives this power-like fall off for distances large
compared to  $M_s$.
}
\item {\emph{Short distances; $r < 1/M_s$}

For smaller $r$-values we find instead that the factor
$e^{-mr}$ is irrelevant, and the result is

\begin{equation}
  \frac{V(r)}{m_1 m_2}=\frac{-1}{2M_D^{n+2}} \frac{S_n}{4\pi}
  \frac{1}{r} \int_0^\infty dm\, m^{n-1}\, e^{-m^2/M_s^2}
  =
  \frac{-\pi^{n/2}}{8\pi }
  \frac{M_s^n}{M_D^{n+2}}\cdot \frac{1}{r}.
  \label{eq:Vsmallr}
\end{equation}

Due to the cut-off, the integral in \eqref{eq:V}
is dominated by $m$-values close to $M_s$ for all $r$-values smaller than
$1/M_s$. Thus, when the distance is smaller than the brane width, the result 
is a potential proportional to $1/r$, corresponding to a standard 
3-dimensional Coulomb potential, although with 
a coupling constant $\sim M_s^n/M_D^{n+2} \sim (M_s R)^n G_{N(4)}$ 
instead of $G_{N(4)}$. Thus the coupling is enhanced by a factor
$\sim(M_s R)^n = (\frac{\rm{compactification \ radius}}{\rm{brane \ width}})^n$.
}

\end{itemize}

\subsection{Eikonal}

In a similar way we can calculate the eikonal $\chi(b)$ by a two-dimensional 
Fourier transform in the transverse coordinates:

\begin{eqnarray}
  \chi(b) &=& \frac{1}{2s} \int \frac{d^2 \kperp}{(2\pi)^2}\,
  e^{i\kperp \bperp} A_{\rm{Born}}(-\kperp^2) =\nonumber \\
  &=& \frac{-s}{2 M_D^{n+2}} \frac{S_n}{(2\pi)^2}
  \int_0^\infty dm\, m^{n-1}\, e^{-m^2/M_s^2} \int d^2 \kperp 
  e^{i \kperp \bperp}
  \frac{1}{m^2 + \kperp^2} = \nonumber \\
  &=&   \frac{-s}{2M_D^{n+2}} \frac{S_n}{(2\pi)^2} 
  2 \pi \int_0^\infty dm\, m^{n-1} e^{-m^2/M_s^2} \int_0^\infty
  \frac{k_\perp d k_\perp}{m^2 + k_\perp^2} \,J_0(k_\perp b) = 
  \nonumber \\
  &=&   \frac{-s}{2M_D^{n+2}} \frac{S_n}{2\pi}
  \int_0^\infty dm\, m^{n-1} e^{-m^2/M_s^2} K_0(m b).
  \label{eq:chiv1}
\end{eqnarray} 

This integral can be expressed in terms of confluent hypergeometric functions 
of the second kind:

\begin{eqnarray}
  \label{eq:chi}
  \chi(b)=-\frac{s M_s^n}{M_D^{n+2}}\Gamma(\frac{n}{2})\frac{\pi^{n/2-1}}{8}
  U(\frac{n}{2},1,\frac{M_s^2 b^2}{4}).
\end{eqnarray}
This expression can easily be used in numerical calculations. For
an intuitive picture, the result for large and small $b$-values can be 
estimated in the same way as the approximations in 
eqs. (\ref{eq:Vbigr}, \ref{eq:Vsmallr}).

\begin{itemize}

\item {\emph{Large impact parameters; $ b \gg 1/M_s$}

For large arguments the asymptotic behavior of the 
Bessel function $K_0(mb)$ is proportional to $\exp(-mb)/\sqrt{mb}$. 
This implies that for large $b$ the Gaussian cut-off
is unessential, and we find the eikonal

\begin{eqnarray}
  \chi(b) &\approx& 
  \frac{-s}{2M_D^{n+2}} \frac{S_n}{2\pi}
  \int dm \, m^{n-1} \, K_0(mb) =
   \frac{-s}{M_D^{n+2}} \frac{S_n}{\pi} 
   2^{n-4}\,\Gamma^2 (n/2)\cdot\frac{1}{b^n} \nonumber \\
  &=&\frac{-s}{2M_D^{n+2}} (4\pi)^{\frac{n}{2}-1} 
   \Gamma\left(\frac{n}{2}\right)\cdot \frac{1}{b^n}.
  \label{eq:chibigb}
\end{eqnarray}
}
\item {\emph{Small impact parameters;  $b \ll 1/M_s$}

For small arguments we have $K_0(mb)\approx \ln(1/(mb))$, which implies

\begin{eqnarray}
  \chi(b) 
  &\approx& 
  \frac{-s}{2M_D^{n+2}} \frac{S_n}{2\pi}
  \int dm \, m^{n-1} \, e^{-m^2/M_s^2} \ln\left(\frac{1}{mb}\right)
  \nonumber\\
  &=&
  \frac{\pi^{\frac{n}{2}-1}}{4} \frac{s}{M_D^2} \left(\frac{M_s}{M_D}\right)^n
  \left( \ln(M_s b) +\frac{1}{2} \psi(\frac{n}{2}) \right)
  \label{eq:chismallb}
\end{eqnarray}
where $\psi(\frac{n}{2})$ is the psi or digamma function.
}
\end{itemize}

Thus we see that the eikonal falls off $\propto 1/b^n$ for large $b$,
and grows logarithmically when $b \rightarrow 0$.
Using the quantity $b_c$ from \eqref{eq:chibc} and keeping only 
the dominant term $\ln(M_sb)$
in \eqref{eq:chismallb}, we can write the results in the form

\begin{eqnarray}
  \chi(b) & \approx & -\left(\frac{b_c}{b}\right)^n; \,\,\, b>b_d  
  \label{eq:Xlarge}  \\
  \chi(b) & \approx & \frac{-(b_c\,M_s)^n}{2^{n-1} \Gamma(n/2)} 
  \ln \left( \frac{1}{M_s b} \right)
  ; \,\,\, b<b_d
  \label{eq:Xsmall} \\
  \mathrm{with} \,\,\, b_d &\equiv& \frac{1}{M_s}
  \label{eq:bd}\\
  \,\, \mathrm{and}\,\,
  b_c&\equiv&\left[ \frac{s(4\pi)^{\frac{n}{2}-1}\Gamma(n/2)}
    {2 M_D^{n+2}}\right]^{1/n}.
  \label{eq:bc2}
\end{eqnarray}
The separation line $b_d=1/M_s$ is an estimate of the $b$-value where 
$\chi(b)$ changes behavior. 
As an example fig. \ref{fig:X} shows these
approximations for $\chi$ together with the exact result for $n=3$
and $\sqrt{s} = M_D =1$ in units such that $M_s=1$.
As $\chi$ is proportional to $s/M_D^{n+2}$, a change in
$s$ and/or $M_D$ just corresponds to a shift of all curves the same distance
up or down.

\FIGURE[t]{%
    \epsfig{file=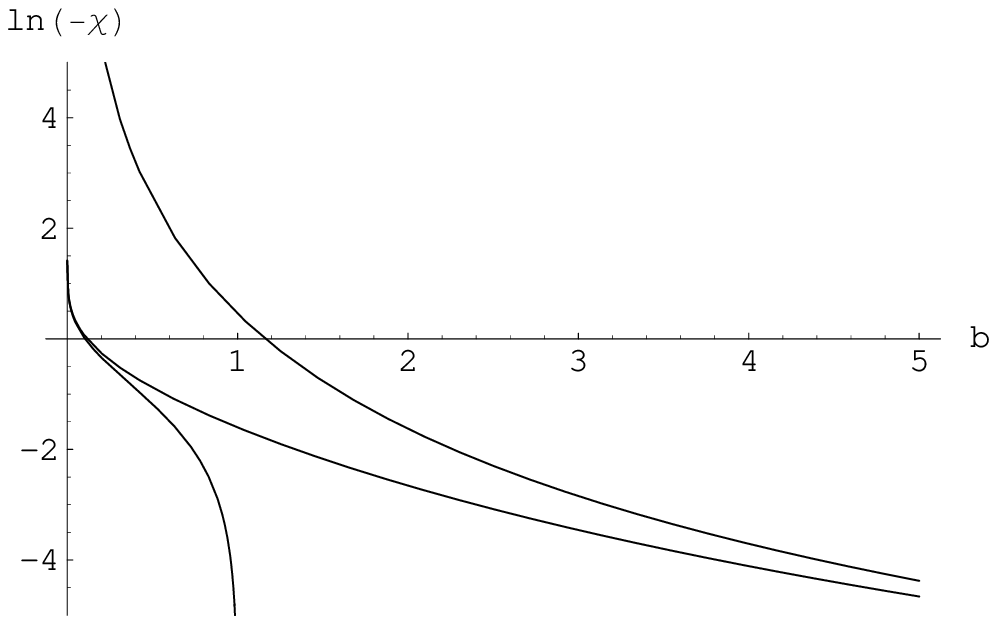,width=10cm}
  \caption{\label{fig:X} The logarithm of $|\chi|$ as a function of 
impact parameter for $n=3$.
The curves show the example where $\sqrt{s}$, $M_D$, and $M_s$ have the 
same magnitude, and the units are chosen such that $\sqrt{s}=M_D=M_s=1$.
This also implies that $b_d=1$. 
The uppermost line is the large $b$ limit of $\chi$ taken from 
\protect\eqref{eq:Xlarge} and the lowermost line is the small $b$ limit 
of $\chi$ taken from \protect\eqref{eq:Xsmall}. The interpolating line is the
exact expression \protect\eqref{eq:chi}. Note that a change in $s$ and/or $M_D$,
keeping $M_s$ constant, just corresponds to shifting all curves up or down.
}
\vspace*{10mm}
}

\section{Higher order loop corrections}
\label{sec:loop}

We note that three different energy scales enter the expressions 
for the Born amplitude in eqs. (\ref{eq:ABorn}, \ref{eq:chi}): 
$\sqrt{s}$, $M_D$, and $M_s$.
Here $\sqrt{s}$ is the total energy in the 
scattering, $M_D$ is the fundamental Planck scale determined by the 
compactification radius $R$, and $M_s$ is related to the width of 
the brane (or the brane tension). 
The result depends on the relative magnitude of these quantities,
and in the following we will successively discuss five different 
kinematical regions, which are illustrated in fig.~\ref{fig:sMs}.

\FIGURE[t]{%
      \epsfig{file=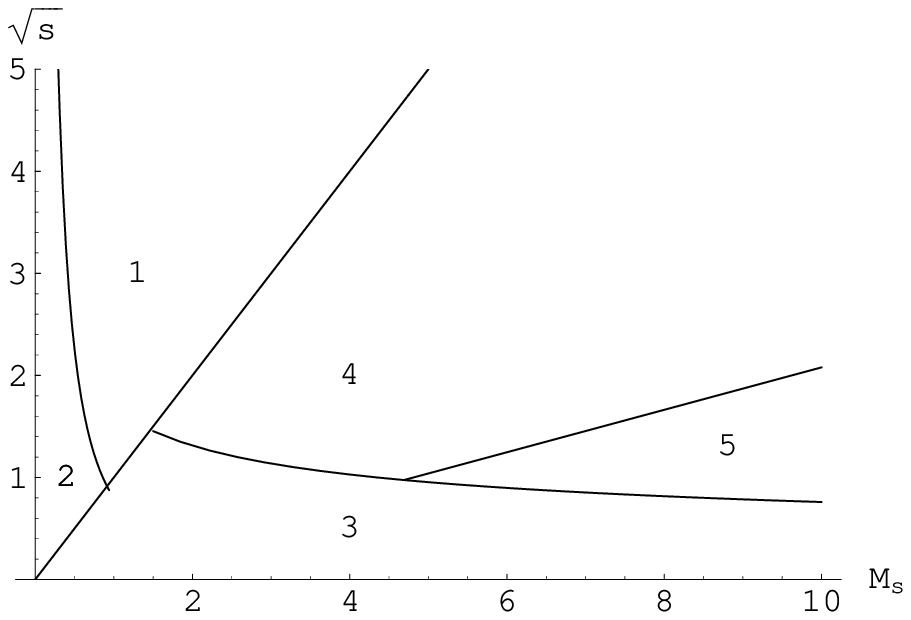,width=10cm}
\caption{\label{fig:sMs} The $(\sqrt{s},M_s)$-plane for $n=3$ and $M_D=1$. 
The straight line separating region 1 and 4 is $\sqrt{s}=M_s$ 
while straight line separating region 4 and 5 is the line where the real 
and imaginary parts in \eqref{eq:1-loop} have equal magnitude.
The power-like curve separating region 1 and 2 is $s_{cd}$ from
\eqref{eq:scd} as a function of $M_s$ and the line separating the regions 
4 and 5 from region 3 is the line where $|A_{\rm{Born}}X|=1$, 
see \eqref{eq:1-loop}. 
In the regions 
1 and 2 $s$ is larger than $M_s$, and, at least for $s \gg M_s$, the 
eikonal approximation is correct. In region 1 the eikonal is, depending 
on $b$, either large compared to 1 or given by \eqref{eq:Xlarge}. 
In region 2 on the other hand 
the $b$-range where $\chi$ is small includes a region where it is
described by \eqref{eq:Xsmall}. In region 3 the correction 
corresponding to higher order loops is small, but in region 4 it is 
important to assure unitarity. 
} }

\subsection{Eikonal regions, $s \gg M_s^2$}
\label{sec:eikonal}

We study the scattering process in the overall cm system, where the 
momentum exchange has no 0-component, $k=(0;\bar{k})$ and $t=-\bar{k}^2$.
From \eqref{eq:ABornbigt} we see that $A_{\rm{Born}}$ falls off 
$\propto 1/\bar{k}^2$ for $\bar{k}^2 > M_s^2$. 
Thus for high energies, such that $s \gg M_s^2$, corresponding to 
region $1$ and $2$ in fig. \ref{fig:sMs}, the
Born term is dominated by small values of $\bar{k}^2 /s$, i.e. small angles.
This implies that the eikonal approximation is applicable. We note
in particular that it is $M_s$ rather than $M_D$, that sets the scale for 
when the eikonal approximation is relevant. The one-loop contribution is 
here dominated by its imaginary part, obtained when the particles 
in the intermediate state $i$ in fig.~\ref{fig:Highs}a are on shell. The
contributions from multi-loop ladder diagrams (fig.~\ref{fig:Highs}b) 
exponentiate, and the scattering 
amplitude is given by \eqref{eq:Aeik}:
\begin{equation}
  \label{eq:Aeik2}
  A_{\rm{eik}}(k^2)
  =2 i s \int d^2 \bperp e^{i \kperp\cdot \bperp}(1-e^{i\chi}).
\end{equation}

From \eqref{eq:Aeik2} we see that the higher order corrections are
important when $\chi$ is of order 1 or larger. Correspondingly the Born term
dominates when $|\chi|<1$. We see from 
eqs.~(\ref{eq:Xsmall}, \ref{eq:bd}) that $\chi$ varies
only logarithmically when $b$ is decreased below the point $b=b_d$.
The importance of the higher order corrections for the integrated cross section
therefore depends on
whether or not $|\chi(b_d)| > 1$. This relation is satisfied whenever
$b_c>b_d$, or equivalently when $s>s_{cd}$, with $s_{cd}$ given by
\begin{equation}
  s_{cd} = \frac{2}{(4\pi)^{\frac{n}{2}-1}\Gamma(\frac{n}{2})} 
  \frac{M_D^{n+2}}{M_s^n}.
  \label{eq:scd}
\end{equation}
This defines the boundary between region 1 and region 2 in fig.~\ref{fig:sMs}. 
In region 1 higher order terms are important for $b<b_c$, and the 
exponentiation in \eqref{eq:Aeik2} is essential to keep the amplitude
within the unitarity constraints.

The difference between regions
1 and 2 is illustrated in fig.~\ref{fig:2ex}. Fig.~\ref{fig:2ex}a corresponds
to region 1, where the energy is high, and $b_c>b_d$. The absolute value of 
the eikonal $\chi$ is smaller than 1 for $b>b_c$, and in this range the 
approximation in \eqref{eq:Xlarge} is relevant. 
For $b<b_c$, $|\chi|$ is large and
rapidly varying, which causes the exponent in \eqref{eq:Aeik2} 
to oscillate rapidly. 

Fig.~\ref{fig:2ex}b corresponds
to region 2. Here $|\chi|<1$ except in a very small region

\begin{equation}
  b < \frac{1}{M_s}
  \exp\left(- \frac{4 M_D^{n+2} \pi^{1-\frac{n}{2}}}{s M_s^n}  \right)
  \label{eq:bl}
\end{equation}
around the origin.
Therefore the Born term dominates the cross section, and
higher order terms give only small corrections.

\FIGURE[t]{%
      \epsfig{file=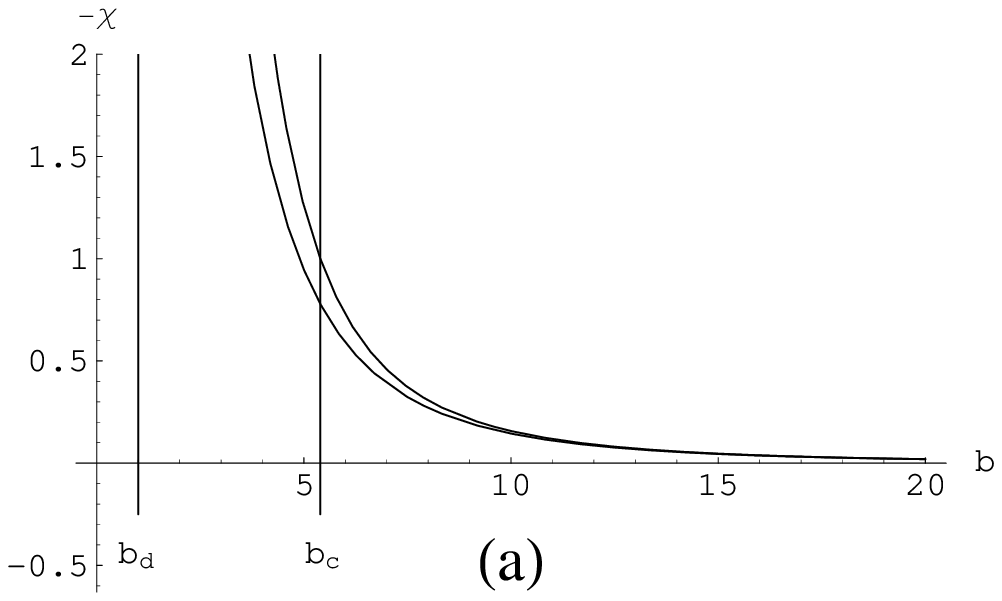,width=7cm}\hfill
      \epsfig{file=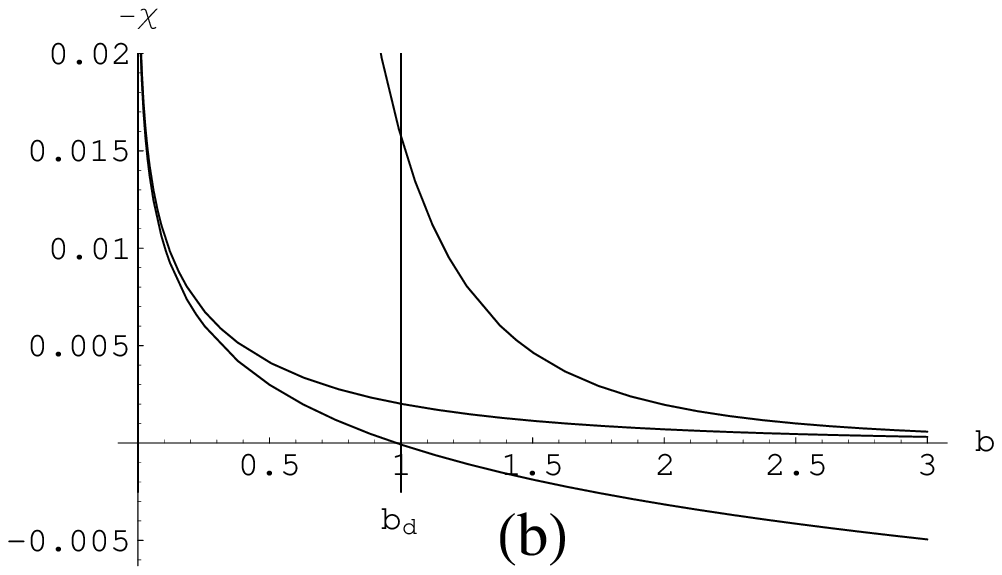,width=7cm}
\caption{\label{fig:2ex} $-\chi$ as a function of 
impact parameter for two examples with $n=3$. 
(a) High energies corresponding to region 1 in fig.~\ref{fig:sMs}, 
with $\sqrt{s}=10$ TeV, $M_s=1$ TeV and $M_D=1$ TeV. 
The upper curve is the approximate expression in \eqref{eq:chibigb},
and the lower curve the exact expression \eqref{eq:X}.
(b) Kinematics corresponding to region 2 in fig.~\ref{fig:sMs}, 
$\sqrt{s}=0.1$ TeV $M_s=1$ TeV and $M_D=1$ TeV.
The upper curve is the approximate high $b$ expression in \eqref{eq:Xlarge},
the lower curve the approximate low  $b$ expression in \eqref{eq:Xsmall}
and the interpolating line is the exact expression in \eqref{eq:chi}.}}

\FIGURE[t]{%
      \epsfig{file=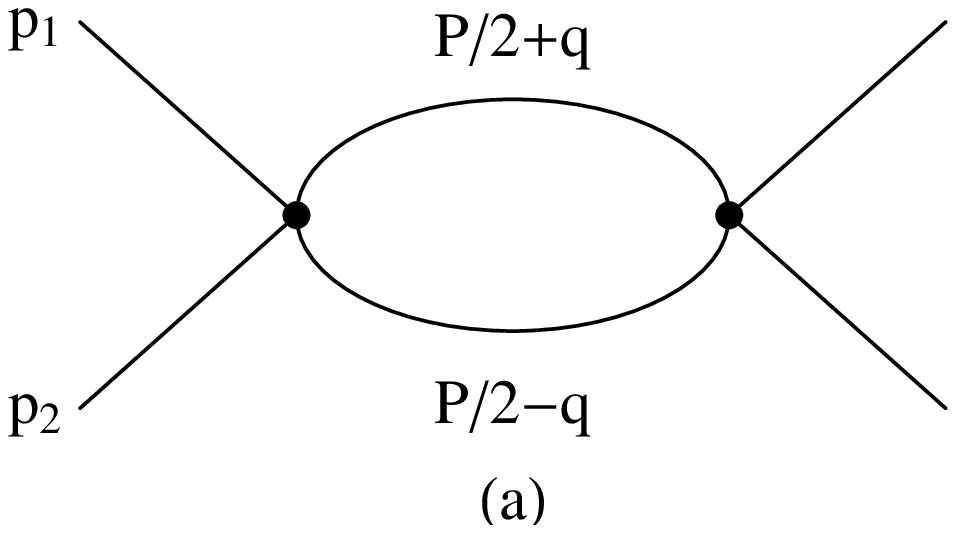,width=7cm}\hfill
      \epsfig{file=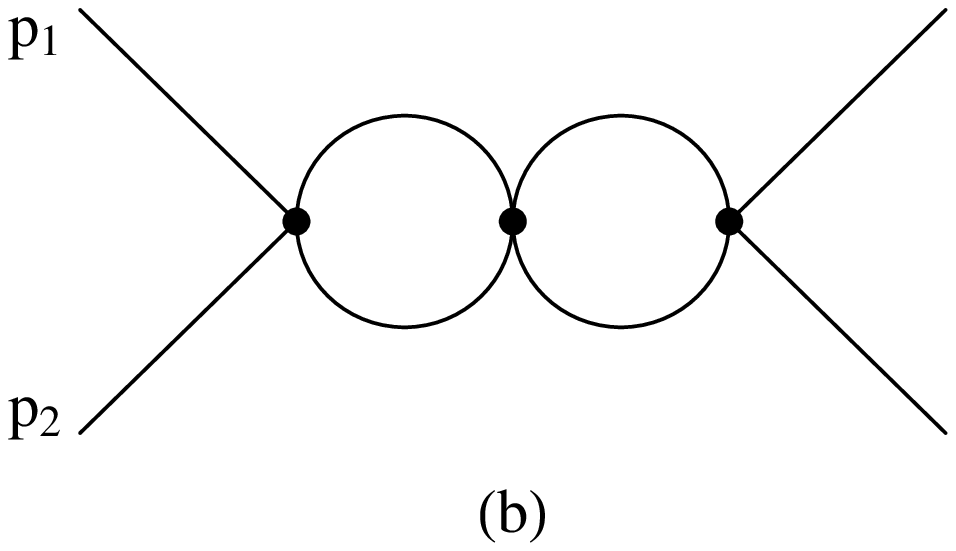,width=7cm}
\caption{\label{fig:Lows} 
When the exchanged momentum is small compared to $M_s$,
the KK propagators are effectively replaced by vertex factors.
The diagrams in fig.\ \ref{fig:Highs} can then be drawn as above
with only standard model particle lines. 
}}

\subsection{Non-eikonal regions, $s < M_s^2$}
\label{sec:noneikonal}

The Born amplitude in \eqref{eq:ABornsmallt} is almost independent 
of the momentum exchange $\bar{k}$ when $k \ll M_s$. 
When $\sqrt{s}\ll M_s$ (regions $3$, $4$, and 5 in fig. \ref{fig:sMs}) 
this includes all kinematically allowed $\bar{k}$-values, which implies that
the scattering is almost isotropic. Thus the exchange of the KK modes 
corresponds effectively to a contact interaction. 
(For wide-angle scattering we also expect corrections from spin 
polarization. This effect is neglected in the following.)
The one-loop contribution in fig.\ \ref{fig:Highs}a
 is then represented by the diagram in fig.\ \ref{fig:Lows}a, which is easily 
calculated. We denote the momenta in the intermediate state $P/2 \pm q$,
with $P=p_1+p_2$, as indicated in fig.\ \ref{fig:Lows}a, and in the cms we 
have $P=(W, \bar{0})$. The vertices are then given by the Born term in 
\eqref{eq:ABornsmallt}, with an effective cut-off when $q=M_s$. The result 
is therefore

\begin{eqnarray}
  \label{eq:1-loop}
  A_{\rm{1-loop}}(k^2) &=& \frac{-i}{2} 
  \int_{k<M_s} \frac{d^4 q}{(2\pi)^4} A_{\rm{Born}}^2 
  \frac{1}{(P/2-q)^2} \frac{1}{(P/2+q)^2}  = \nonumber \\
  &\equiv& A_{\rm{Born}}^2\cdot X, \,\,\,\,
  \mathrm{with} \,\, 
  X \approx  \frac{1}{32 \pi^2} ( \ln\frac{4M_s^2}{s} + i\pi) 
  \label{eq:X}
\end{eqnarray}
We note here in particular that the result is a constant, independent
of the momentum transfer $k$. Therefore also the one-loop
amplitude can be effectively regarded as a contact term with a cut-off
when $k>M_s$. Consequently the two-loop diagram in fig.~\ref{fig:Lows}b
can be calculated in the same way as the one-loop diagram, and
the result is
\begin{equation}
  A_{\rm{2-loop}} = A_{\rm{1-loop}} \cdot A_{\rm{Born}} X = A_{\rm{Born}}\cdot 
  (A_{\rm{Born}} X)^2.
  \label{eq:X2-loop}
\end{equation}
In the same way we can calculate ladder diagrams with more loops.
Summing all ladders we obtain
\begin{equation}
  A_{\mathrm{ladder} } = A_{\rm{Born}}\, (1+A_{\rm{Born}}X+
(A_{\rm{Born} }X)^2 + \ldots\ ) = \frac{ A_{\rm{Born} } }{ 1-A_{\rm{Born} }X}.
  \label{eq:Xall-loop}
\end{equation}
We see that instead of the exponent in the eikonal regime (where forward 
scattering dominates) we here obtain a geometric series from ladder
type contributions.
The importance of higher order corrections is now determined 
by the quantity $A_{\rm{Born}}\, X$. When $|A_{\rm{Born}}\, X| \ll 1$
the Born term dominates. This corresponds to region 3 in 
fig.~\ref{fig:sMs}. 

When instead $|A_{\rm{Born}}\, X| > 1$, we expect different
results depending on whether it is the real or the imaginary part
which dominates. When $\ln(4 M_s^2/s) < \pi$, the imaginary
part dominates the loop integral in \eqref{eq:1-loop}. Thus this diagram
is dominated by real intermediate states $i$ in fig.~\ref{fig:Highs}a,
and the ladder diagrams in fig.~\ref{fig:Highs}b or
fig.~\ref{fig:Lows}b should be important higher order corrections. 
This corresponds to region 4 in fig.~\ref{fig:sMs}.

When $\ln(4 M_s^2/s) > \pi$ (region 5 in fig.~\ref{fig:sMs})
the real part dominates the loop integral.
This implies that inelastic scattering and virtual intermediate 
states are essential. We then expect important contributions from
more complicated, non-ladder, diagrams, like the examples shown 
in fig. \ref{fig:FD}. For this reason we do not expect the 
result in \eqref{eq:Xall-loop} to be representative for a sum of all
higher order corrections in this kinematical region.

\FIGURE[t]{%
      \epsfig{file=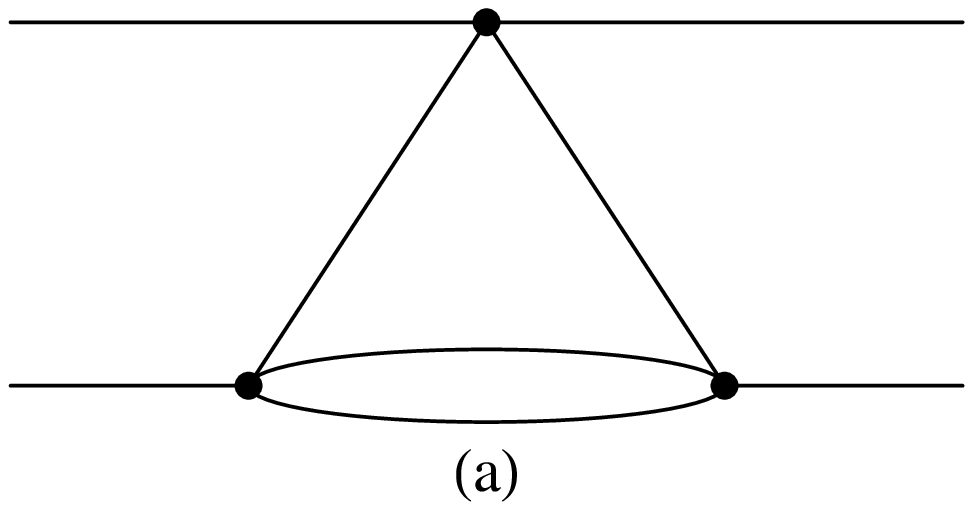,width=7cm}\hfill
      \epsfig{file=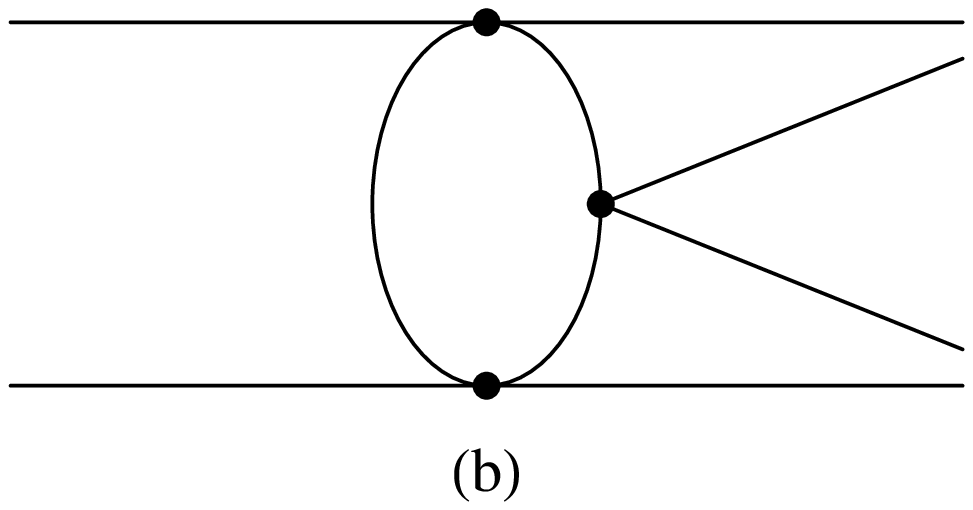,width=7cm}
\caption{\label{fig:FD}(a) An example of a non-ladder diagram 
contributing to the
elastic cross section in region 5 in fig.~\ref{fig:sMs}.
(b)  An example of a diagram contributing to the
inelastic cross section in region 5 in fig.~\ref{fig:sMs}.}}

\section{Cross sections}
\label{sec:cross}
Below we successively discuss the cross sections obtained in the 
five different regions in fig.~\ref{fig:sMs}.

\subsection{Region 1, $s>M_s^2$ and $|\chi(b_d)|>1$}

In this region $s>M_s^2$ and $\chi(b_d)>1$. As discussed in section 
\ref{sec:eikonal} the scattering is suppressed for  $-t>M_s^2$. 
The first constraint therefore means that the cross section 
is dominated by small angle scattering, the imaginary 
part dominates the one-loop contribution, and the eikonal $\chi(b)$ 
exponentiates. The cross section is then given by
\begin{equation}
  \sigma = \int d^2 \bperp \, 2 Re(1 - e^{i\chi(\bperp)}).
  \label{eq:sigmaeikonal2}
\end{equation}
The effect of the constraint $|\chi(b_d)|>1$ was
illustrated in fig.~\ref{fig:2ex}a. It implies that $b_d<b_c$, and that the
approximation $\chi\approx -(b_c/b)^n$ in \eqref{eq:Xlarge} is relevant 
for all $b>b_c$. In particular this means that, for $b>b_c\,(>b_d)$, we have 
$|\chi|<1$ and $2 Re(1 - e^{i\chi(\bperp)}) \approx \chi^2$.  For central
collisions, where $b<b_c$, higher order loop corrections are important 
to satisfy unitarity. Here $|\chi|$ is larger than 1 and rapidly varying, the 
exponent in \eqref{eq:sigmaeikonal2} is oscillating, and therefore
\begin{equation}
  \left< \, 2 Re(1 - e^{i\chi(\bperp)}) \,\right> \,\approx 2.
\label{eq:average}
\end{equation}
(For $b<b_d$ the variation in $\chi$ is only logarithmic, and 
therefore small in \emph{relative} magnitude. As $\chi$ is very large, 
the variation may still be large in \emph{absolute} magnitude and
such that \eqref{eq:average} is valid also here.)
Inserting these results into \eqref{eq:sigmaeikonal2}, we get (for
$n\ge 2$) the following result for the total cross section:
\begin{eqnarray}
  \sigma \approx \int_0^{b_c} d^2 b \cdot 2
  + \int_{b_c}^\infty d^2 b \left(\frac{b_c}{b}\right)^{2n} 
  =\pi b_c^2 (2+\frac{1}{n-1}) = 2\pi \, b_c^2 \,\frac{n-1/2}{n-1}.
  \label{eq:sigmahigh}
\end{eqnarray}
When $s$ is increased, $\sigma$ grows proportional to $b_c^2 \propto s^{2/n}$. 
We note that the cross section is dominated by central collisions with
$b<b_c$
(especially for large $n$), with only a small contribution from larger
impact parameters. 
Integrating the constant 1 in the parenthesis in \eqref{eq:Aeik2} between
0 and $b_c$ gives a dominant forward peak, with oscillations at larger angles. 
The amplitude for these oscillations falls off proportional to $1/k^{3/2}$, 
corresponding to $d\sigma/dt \propto 1/k^{3}$ for
the cross section.

For large $k$ the dominant contribution in \eqref{eq:Aeik2} comes from 
the term $e^{i\chi}$ and a small range of $b$-values around $b_s$, where
\begin{equation}
  \label{eq:bs}
  b_s=b_c\left(\frac{n}{k b_c}\right)^{\frac{1}{n+1}}.
\end{equation}
For these $b$-values the frequencies of the exponents 
$e^{i \kbar \cdot \bbar}$ and $e^{i \chi (\bbar)}$ in \eqref{eq:Aeik2}  
oscillate in phase, which gives an enhanced contribution.
Using the saddle-point approximation we get from this contribution  
(apart from logarithmic corrections) a cross section which falls off 
like $d\sigma/dt \propto 1/t^{\frac{n+2}{n+1}}$. 
This contribution is dominating for $k > n/b_c$, where $|\chi(b_s)|>1$.
As pointed out in \cite{Giudice:2001ce} it corresponds to classical
scattering in a  $1/r^{n+1}$-potential. 
For small scattering angles $\theta$ we have
for a non-relativistic particle with mass $m_1$ moving with constant 
speed $v$ and momentum  $p=m_1 v$ in 
the potential of a mass $m_2$ 

\begin{eqnarray}
  \theta \approx \frac{|\pperp| }{|\pbar|}
  &=& \frac{1}{|\pbar|} \int_{-\infty}^{\infty} dt F_{\perp}(r)  \\ \nonumber
  &=& -G_{N(4)} R^n S_n \Gamma(n) \frac{m_1 m_2}{m_1 v}
  \int_{-\infty}^{\infty} \frac{dr}{v} \frac{d}{db}
  \left( \frac{1}{\sqrt{r^2+b^2}}\right)^{n+1} \\ \nonumber
  &=&\frac{n(2\sqrt{\pi})^{n}\Gamma(\frac{n}{2})}{8 \pi v^2}
  \frac{m_2}{M_D^{n+2}}\frac{1}{b^{n+1}}.
  \label{eq:theta}
\end{eqnarray}
From this we see that if $m_1=s/(4 m_2)$ 
\begin{equation}
  b_{\rm{nonrel}}=b_c\left( \frac{n}{4 v p b_c} \right)^{\frac{1}{n+1}}
  \label{eq:nonrel}
\end{equation}
agreeing parametrically with \eqref{eq:bs}. (A numerical difference is 
expected since \eqref{eq:bs} is ultra-relativistic whereas \eqref{eq:nonrel} 
is a non-relativistic result.)
This behavior is discussed in more detail in \cite{Giudice:2001ce}, and
we note that in this region, where $s$ is much larger than both $M_s^2$
and $s_{cd}$, our result is consistent with the result of this reference.
A necessary condition is, however, that $\sqrt{s}$, $M_D$, and $M_s$
have values such that $b_s> b_d=1/M_s $, which for fixed $k$-value
gives a minimum value for $M_s$. If this relation is not satisfied,
the phase variation in $\exp(i \chi)$ is given by \eqref{eq:Xsmall}
rather than \eqref{eq:Xlarge}, and therefore we do not get
the phase coherence in the integral in \eqref{eq:Aeik2}.

\subsection{Region 2, $s>M_s^2$ and $|\chi(b_d)|<1$}

In region 2, $s$ is larger than $M_s^2$ but smaller than $s_{cd}$,
and therefore $b_d>b_c$. A typical example is illustrated in 
fig.~\ref{fig:2ex}b. We see here that $|\chi|$ is small compared to 1,
apart from the logarithmic peak for very small $b$. 
The influence of the small $b$ peak is also suppressed by 
a phase space factor proportional to $b \,db$. 
The cross section is therefore well approximated by the Born amplitude.

The largest contributions to the cross section come from $b$-values
in the neighborhood of $b_d$; for larger $b$, $\chi$ falls off 
$\propto (b_c/b)^n$, and for smaller $b$ the scattering is limited
by the smaller phase space $\sim b \,db$. These $b$-values are
just in the transition region between the two asymptotic forms in 
eqs.~(\ref{eq:Xlarge}, \ref{eq:Xsmall}). To get a good estimate of the
cross section we should therefore use the exact expression for $\chi$
in \eqref{eq:chi}. For an order of magnitude estimate we may, however, 
approximate $\chi$ by the asymptotic result $\chi\approx -(b_c/b)^n$ for 
$b>b_d$, and by a constant $=-(b_c/b_d)^n$ for all $b<b_d$. This gives the 
following qualitative estimate for the total cross section:
\begin{eqnarray}
  \sigma 
  \sim 
  \int_0^{b_d} d^2 b \left(\frac{b_c}{b_d}\right)^{2n}
  + \int_{b_d}^\infty d^2 b \left(\frac{b_c}{b}\right)^{2n} 
  = \pi \frac{b_c^{2n}} {b_d^{2n-2}}\,\frac{n}{n-1}.
  \label{eq:sigmalow}
\end{eqnarray}

As $b_c \sim (s/M_D^{n+2})^{1/n}$, and $b_d = 1/M_s$,
we note that the cross section grows $\propto s^2 M_s^{2n-2}/M_D^{2n+4}$.
Thus, although the cross section is comparatively small in this region, it
has a stronger growth rate $\propto s^2$ than in region 1. 

For the differential cross section, we note that the t-channel Born amplitude
is proportional to $1/k^2$ for $-k^2\gg M_s^2$. This implies that the 
cross section has a forward peak. It corresponds to 
scattering at distances small compared to $1/M_s$, in the $1/r$ potential 
from \eqref{eq:Vsmallr}.  
There is however no forward divergence since the growth is softened at 
$-k^2 \sim M_s^2$, i.e. at distances comparable to the brane thickness. 

\subsection{Region 3, $s<M_s^2$ and $|\Aborn X|<1$}
\label{sec:Region3}
In region 3 the cross section is also dominated by the Born amplitude.
But in this case the scattering is almost isotropic
(apart from spin dependences) as the factor $-k^2$ in the propagator is 
small compared to the heavier and most important KK modes. 
This implies that we may also have important contributions from 
u- and s-channel exchanges. For identical particles, the u-channel
contribution has the same magnitude as that from t-channel.

\subsection{Region 4, $s<M_s^2$, $|\Aborn X|>1$ and $\mbox{Im}(X)>\mbox{Re}(X)$}
\label{sec:Region4}
The one-loop t-type contribution in fig.~\ref{fig:Lows}a, is 
dominated by the imaginary part, originating from real intermediate states.
If loop diagrams of this type dominate, the all-loop 
amplitude is approximated by the geometric sum in \eqref{eq:X}. 
As in region 3, the result is then approximately isotropic, 
but here higher order corrections give some
suppression compared to the Born approximation.
For identical particles the u-type ladder is identical to the t-type ladder 
and hence equally important.
For particle-antiparticle scattering s-channel contributions have to be
considered.

\subsection{Region 5, $s<M_s^2$, $|\Aborn X|>1$ and $\mbox{Im}(X)<\mbox{Re}(X)$}
\label{sec:Region5}
In this region the one-loop diagram has a dominant real part. This implies
that virtual intermediate states and inelastic reactions are important.
Therefore non-ladder diagrams are expected to give large contributions,
and we showed two examples in fig. \ref{fig:FD}. 
This region is consequently much more
complicated than the other kinematical regions. It corresponds
to situations where the effective cut-off $M_s$ is large ("narrow brane" 
or strong "brane tension") and the energy is in an intermediate range.
From fig.\ \ref{fig:sMs} we see that for, e.g. $n=3$, $M_s$ must be larger 
than 5 $M_D$. 
In this paper we will not make any specific predictions for what might 
be expected in this kinematic region.

\section{Conclusions}
\label{sec:Conclusion}

In the ADD model it is assumed that standard model particles live on a 
4-dimensional brane, embedded in a $(4+n)$-dimensional space with
$n$ compactified dimensions. In these only the gravitational field is allowed
to propagate. If the brane is infinitely thin and infinitely rigid, 
the exchange of very massive Kaluza--Klein modes represents a contact 
interaction of infinite strength between the standard model particles. 
This is not physically acceptable and different ideas have been proposed 
to regularize the scattering process. 

If the brane has a finite width, or if it is not infinitely well 
localized, the exchange of KK modes will be suppressed for 
KK wavelengths shorter than the 
width of the brane, or the size of its fluctuations. 
This will therefore give an effective cut-off (denoted $M_s$) for high 
KK masses, which does not have to be of the same magnitude as the 
fundamental Planck mass $M_D$.

In this paper we have studied the effect of such a cut-off on the 
scattering of standard model particles at various energies. We find that 
several troublesome infinities and divergencies are removed. The scattering
process depends on three different energy scales, the collision energy
$\sqrt{s}$, the fundamental Planck scale $M_D$, and the cut-off scale
$M_s$. The Planck scale, $M_D=(8\pi R^n G_{N(4})^{-1/(n+2)}$, depends on the 
compactification radius $R$ of the extra dimensions and the magnitude of 
Newton's constant, while the effective cut-off depends on the width of the 
brane, $M_s \sim(\mathrm{brane}\ \mathrm{thickness})^{-1}$, 
or the fluctuations in its position. 
These scales are thus not automatically related.
Clearly the compactification scale $R$ must be larger 
than the brane width $1/M_s$.

Depending on the relative magnitude between these scales,
we have here studied five different kinematical regions
with different dynamical behavior. 
In one region (region 1 in fig.~\ref{fig:sMs}), the scattering
is dominated by small angles, and the eikonal approximation is applicable.
Here we recognize classical scattering in a $1/r^{n+1}$ potential and
the results of Giudice-Rattazzi-Wells \cite{Giudice:2001ce}. 
In two other regions (2 and 3 in fig.~\ref{fig:sMs}) the Born approximation 
is applicable. In one of these (region 2) forward scattering dominates,
and corresponds to scattering in a $1/r$ potential, but with a coupling
enhanced by a factor proportional to 
$(\frac{\rm{compactification \ radius}}{\rm{brane \ width}})^n$
compared to scattering in the ordinary $1/r$ Newtonian large distance
potential. In the other Born region (region 3) the scattering is 
approximately isotropic, as expected in \cite{Han:1998sg,Giudice:1998ck}.
In a fourth region the exponentiation from ladder-type diagrams 
in the eikonal region is replaced by a geometric sum.
The scattering is expected to be mostly elastic since on-shell intermediate
states dominate, but approximately isotropic. 
In the last region inelastic processes and non-ladder loop diagrams are
important and make predictions very difficult. 
The boundaries between the different regions are expressed in the 
three mass scales involved, as illustrated in fig.~\ref{fig:sMs}.

\section*{Acknowledgments}
We thank Leif Lönnblad and Johan Bijnens for useful discussions.

\bibliographystyle{utcaps}  
\bibliography{references,refs}
\end{document}